\documentclass[aps,prl,twocolumn,amssymb,superscriptaddress]{revtex4}
\usepackage{subfigure}
\usepackage{graphicx}
\usepackage{rotating} \usepackage{color}
\usepackage{amsmath,amsthm}
\usepackage{epstopdf}
\begin{document}

\title{Confinement-induced glassy dynamics in a model for chromosome organization}



\author{Hongsuk Kang}
\affiliation{Institute for Physical Sciences and Technology, University of Maryland, College Park, MD 20742}

\author{Young-Gui Yoon}

\affiliation{Department of Physics, Chung-Ang University, Seoul 156-756}

\author{D. Thirumalai}
\affiliation{Institute for Physical Sciences and Technology, University of Maryland, College Park, MD 20742}

\author{Changbong Hyeon}
\thanks{hyeoncb@kias.re.kr}
\affiliation{Korea Institute for Advanced Study, Seoul 130-722}

\date{\today}

\begin{abstract}
Recent experiments showing scaling of the intrachromosomal contact probability, $P(s)\sim s^{-1}$ with the genomic distance $s$, are interpreted to mean a self-similar fractal-like chromosome organization. However, scaling of $P(s)$ varies across organisms, requiring an explanation. 
We illustrate dynamical arrest in a highly confined space as a discriminating marker for genome organization, by modeling chromosome inside a nucleus as a homopolymer confined to a sphere of varying sizes. 
Brownian dynamics simulations show that the chain dynamics slows down as the polymer volume fraction ($\phi$) inside the confinement approaches a critical value $\phi_c$.
The universal value of $\phi_c^{\infty}\approx 0.44$ for a sufficiently long polymer ($N\gg 1$) allows us to discuss genome dynamics using $\phi$ as a single parameter.
Our study shows that the onset of glassy dynamics is the reason for the segregated chromosome organization in human ($N\approx 3\times 10^9$, $\phi\gtrsim\phi_c^{\infty}$), whereas  chromosomes of budding yeast ($N\approx 10^8$, $\phi<\phi_c^{\infty}$) are equilibrated with no clear signature of such organization. 
\end{abstract}

\maketitle

Chromosomes exhibit dramatic changes in their spatial organization along the cell cycle. 
In the metaphase, they are condensed into compact blob-like structures \cite{nagano2013Nature}, whereas in the interphase they decondense to a less compact coil-like structures.
Interphase chromosomes are not random but form territories \cite{cremer2001NRG}, and their organization may be fractal-like \cite{lieberman09Science}.
Advances in experimental techniques \cite{bolzer2005PLoSBiol,Dekker2002Science, lieberman09Science,Rao2014Cell,dekker2013NRG}
have provided quantitative details of chromosome organization in the form of chromosomal contact maps describing how distant loci are structurally organized.
The contact probability of two loci separated by a genomic distance $s$ scales as $P(s)\sim s^{-1}$, differing from $P(s)\sim s^{-1.5}$ in equilibrated polymer melts. 
The deviation of the exponent from $-1.5$ is taken as an evidence that chromosomes form a non-equilibrium globule with segregated domains rather than a fully equilibrated globule with entanglements \cite{lieberman09Science,Mirny11ChromoRes}. 
Such an interpretation of the structural organization based solely on $P(s)$ is not universally accepted \cite{Barbieri12PNAS}. In addition, genome structure could vary depending on the extent of maturity of human cells \cite{Zhang15PNAS}. 
Still, the scaling of $P(s)$ varies depending on organisms. 
It is therefore important to develop a theoretical framework for distinguishing between genome structure in different organisms.   

From a biological perspective, it could be argued that the hierarchical and scale-free organization of chromosome, without knots, is beneficial for access to a target locus \cite{grosberg1993EPL} or for the faster response to an environmental change by easing the condensation-decondensation process \cite{Rosa2008PLoSCB,Halverson14RPP}.
Although the origin of chromosomal territories is controversial because equilibrium polymer configurations with many loops naturally produce segregated domains as well \cite{Muller2000PRE,mateos2009PNAS}, 
a major non-equilibrium effect, glassy dynamics of the genome under strong confinement, should not be overlooked from a contributing factor in chromosome folding. 
The relaxation time of a polymer via disentanglement \cite{Sikarov94BJ} ($\tau_{rep}\sim N^3$ \cite{deGennesbook,grosberg1988JP,Rosa2008PLoSCB,Halverson14RPP}) could be far longer, effectively permanent for higher organisms, than the cell cycle time ($\tau_{cell}$) \cite{Rosa2008PLoSCB,Halverson14RPP} for a large $N$.  
Furthermore, a substantial increase of polymer relaxation time is also expected in a strong confinement as is the case for DNA inside viral capsid \cite{Berndsen14PNAS} even when $N$ is not too large. 
Thus, to fully describe the genome structuring, it is imperative to understand the polymer dynamics under confinement and how it might vary across various species. 
The major goal of this work is to develop a physical basis, using  relaxation dynamics as a quantitative measure, to discriminate between genome organization in different organisms.  

\begin{figure*}[t]
\includegraphics[width=1.8\columnwidth]{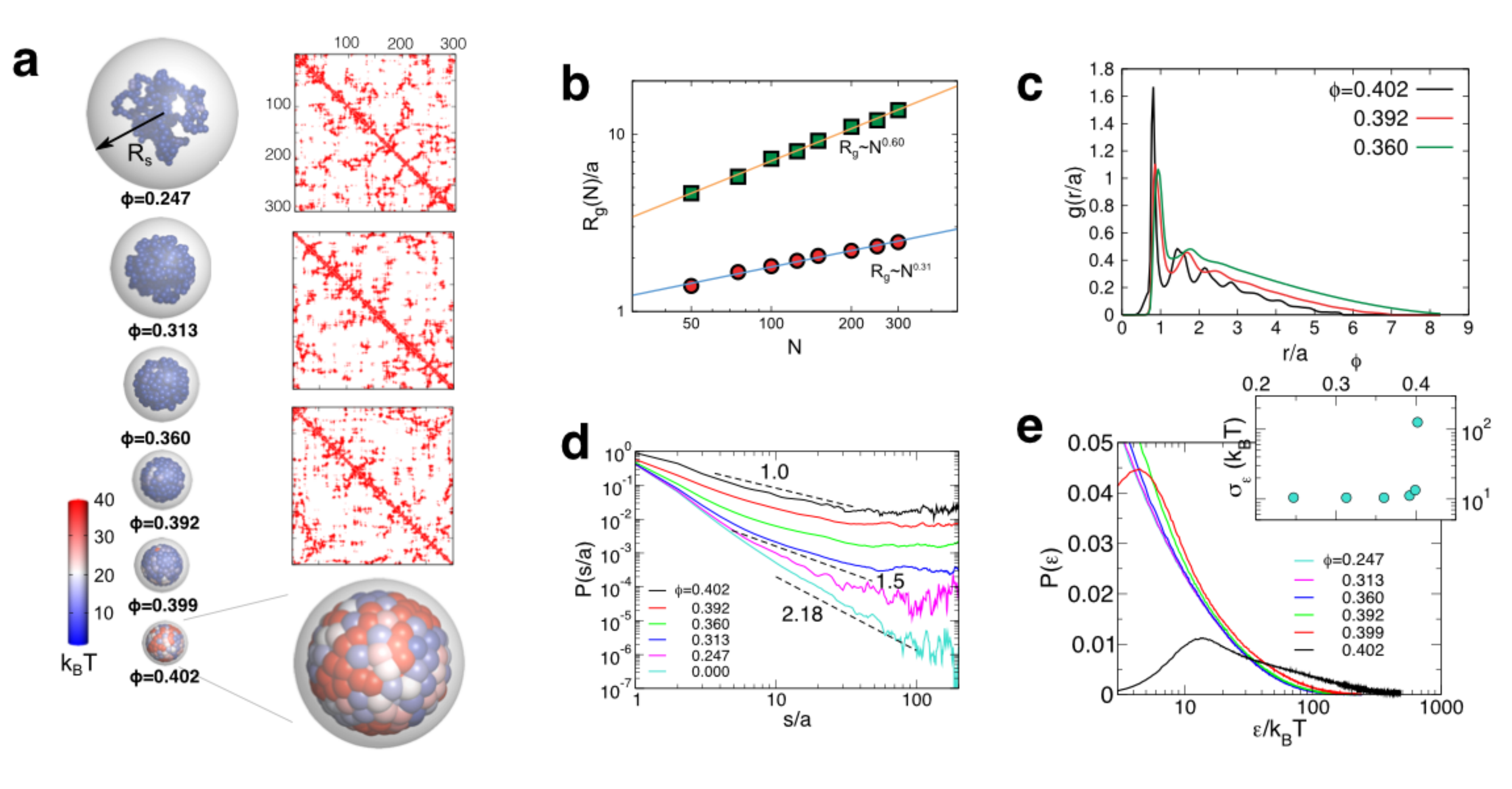}
\caption{Polymer ($N=300$) confined to spheres. 
(a) Snapshots from simulations. 
The value of potential energy (scale on the left) for each monomer shows that 
the spatial heterogeneity increases as $\phi$ approaches $\phi_c(300)\approx 0.404$ (see Fig.S5 for other snapshots).  
Contact maps from three distinct polymer configurations at $\phi=0.402$ near $\phi_c(300)$ are shown on the right. 
(b) Flory laws $R_g\sim N^{3/5}$ and $\sim N^{1/3}$ are satisfied for unconfined and strongly confined ($\phi=0.402$) chains of varying $N$, respectively. 
(c) RDFs (see SM) at three $\phi$ values. 
(d) Inter-segmental contact probabilities (see SM for definition) with  increasing $\phi$ from 0 to 0.404 ($N=300$). 
(e) Distributions of monomer energy ($\epsilon_i=U^{\text{bond}}_i+\sum_{j\neq i}U^{\text{ex}}_{i,j}$, where the surface interaction term is excluded from the calculation. See SM), $P(\epsilon)=N^{-1}\sum_{i=1}^N\delta(\epsilon_i-\epsilon)$, for increasing $\phi$. Divergence of the standard deviation ($\sigma_{\epsilon}^2=\langle \epsilon^2\rangle-\langle \epsilon\rangle^2$ where $\langle \epsilon^n\rangle=\int_0^{\infty}\epsilon^nP(\epsilon)d\epsilon$) near $\phi_c$ is shown in the inset.  
\label{compaction}}
\end{figure*}

Although explicit models that consider circular DNA or multi-chains and specific contacts based on Hi-C contact maps \cite{le2013Science,Tokuda2012BJ,Ganai2014NAR} are possible,
here we study the dynamics of homopolymers confined to a sphere of varying sizes as a first step towards understanding the dynamical features of interphase chromosomes.   
We consider a single self-avoiding polymer chain, representing chromatin fiber, confined to a sphere and employ dynamical measures previously used to study supercooled liquids \cite{kirkpatrick1988PRA,kang2013PRE,Kirkpatrick15RMP} as a vehicle to investigate non-equilibrium effects.  
Our major finding is that the dynamics and organization of homopolymer vary dramatically as the extent of confinement is increased. 
When this result is translated into genome organization, 
we find that bacteria and yeast chromosome folding can be thought of as an equilibrium process whereas glassy behavior governs the territorial organization in humans. 
These inferences cannot be drawn from genome contact maps alone, which has been the sole focus on chromosome folding.

The equilibrium aspects of confined polymers are well understood \cite{deGennesbook,Daoud75Macromolecules}. 
The equilibrium free energy of polymer confined to a sphere is not extensive  \cite{GrosbergBook,Cacciuto2006NanoLett} in contrast to polymer localization in a slit or a cylinder.  
Furthermore, as the extent of confinement increases, the volume fraction, defined by $\phi=(R_g^c/R_s)^3$ (Fig.\ref{compaction}a, see Supplemental Material) increases, and more importantly, the equilibration time of the chain ($\tau_{eq}$) increases dramatically.  
If $\tau_{eq}$ for a genome is longer than finite cell doubling time ($\tau_{cell}$) then the decondensation-condensation cycle dynamics of the genome should be under kinetic control. 
We explore these aspects in the context of genome folding using simulations of homopolymers confined in a sphere (see SM for details),  
highlighting the confinement effect on polymer leading to the ultraslow glassy dynamics, such that $\tau_{eq}\gg\tau_{cell}$, which we will show is the case in human chromosomes ($N\approx 10^9$) and viral DNA ($N\approx 10^5$).

In general, it is difficult to distinguish between non-equilibrium conformation of a polymer from its equilibrium counterpart because 
polymer configurations for both cases could be similar, just as is the case for liquids and glasses. 
Indeed, the polymer size with increasing $N$ satisfies the Flory relationship, $R_g\sim N^{\nu}$ with $\nu\approx 3/5$ and $1/3$ for weak and strong confinement, respectively (Fig.\ref{compaction}b), crossing over the regime $R_g\sim N^{1/2}$ at $\phi^{(\theta)}\approx 0.2$ where repulsion due to excluded volume is counter-balanced by the confinement pressure. Therefore, in strong confinement $R_g$ scaling cannot distinguish between equilibrium and non-equilibrium globules. 
The radial distribution function (RDF) between monomers at high $\phi$ is reminiscent of the closely packed structure (Fig.\ref{compaction}c), suggesting that extent of confinement controls the chain organization. 

The scaling exponent $\alpha$ of the contact probability between two sites separated by the chain contour $s$, $P(s)\sim s^{-\alpha}$, is one way to assess the chain organization (Alternatively, the average distance between two loci separated by $s$, $R(s)\sim s^{\nu}$, can be used \cite{mateos2009PNAS,Barbieri12PNAS}. See Fig.S3). 
$P(s)\sim s^{-2.18}$ is expected for unconfined self-avoiding walk (SAW) (see SM) \cite{desCloizeauxJP80,Redner80JPA,Toan08JPCB}. 
For an equilibrium globule under  strong confinement, polymer chains are  in near $\Theta$-condition because of the effective cancellation between attraction and repulsion. 
Hence, we expect that  $P(s)\sim s^{-\alpha}$ with $\alpha=1.5$ \cite{deGennesbook,Lua2004Polymer}.
In the case of strong confinement, however, $P(s)\sim s^{-1}$ in the range of $s/a\sim \mathcal{O}(10)$ for $N=300$, similar to the scaling observed in the Hi-C analysis of the chromosome in interphase \cite{lieberman09Science,Gursoy2014NAR}. 
The range of $s^{-1}$-scaling increases as the extent of confinement increases (Fig.\ref{compaction}d). 

However, $P(s)$ scaling is not an indicator of the underlying dynamics. 
Even for a SAW chain with no specific attractive interaction, dynamics can be arrested in strong confinement, preventing a full equilibration of the chains on relevant time scales.
We document the emergence of glassy behavior under strong confinement by first calculating the potential energy of each monomer (Figs.\ref{compaction}a, \ref{compaction}e, S5).  
The spatial heterogeneity of monomer energies at $\phi=0.402$ is striking (Figs.\ref{compaction}a, S5), which is also indicated by the abrupt changes in the monomer energy distribution $P(\epsilon)$ (Fig.\ref{compaction}e) and the standard deviation $\sigma_{\epsilon}$ (Fig.\ref{compaction}e, inset). 

\begin{figure}[t]
\includegraphics[width=1.0\columnwidth]{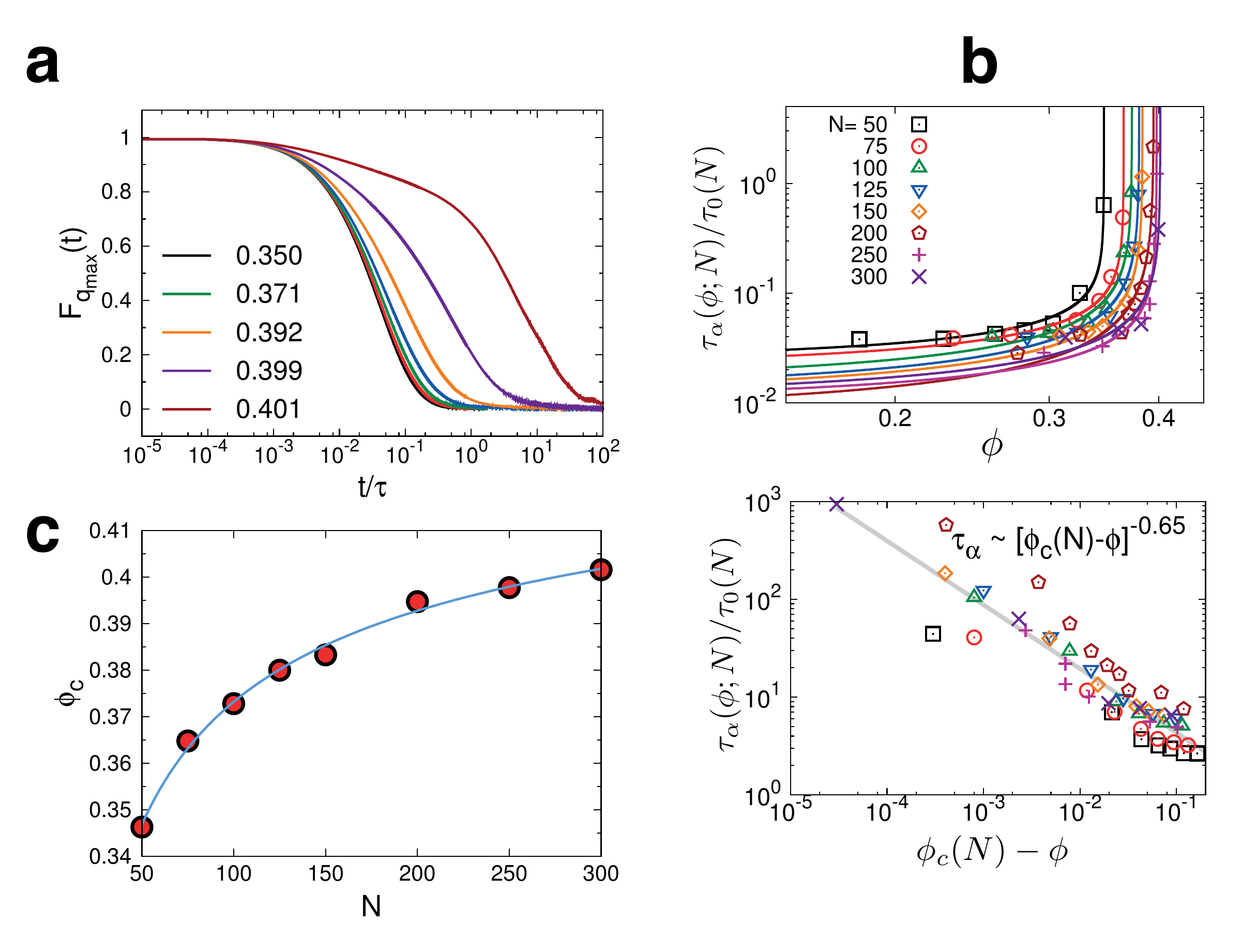}
\caption{Polymer dynamics under confinement probed using $\tau_{\alpha}(\phi)$.
(a) $\langle F_{q_{\text{max}}}(t)\rangle$ with varying $\phi$ for $N=300$.
The time on the abscissa is scaled by $\tau=a^2/D$
(see SM).  
(b) (top) For polymer with $N$, $\tau_{\alpha}(\phi;N)$ are fit to Eq.\ref{eqn:tau_scaling}.
(bottom) To obtain the universal scaling exponent of $\tau_{\alpha}$ near $\phi_c(N)$, 
the fit was made using $\tau_{\alpha}/\tau_0(N)=(\phi_c(N)-\phi)^{-\nu_{\tau}}$ for all $N$, which confers $\nu_{\tau}=0.65$. 
(c) Finite size scaling to obtain $\phi_c^{\infty}\equiv\phi_c(N\gg 1)$.
$\phi_c(N)$s fitted to
$\phi_c(N)=\phi_c^{\infty}-aN^{-\nu'}$ give 
$\phi_c^{\infty}=0.449$, $a=0.552$, and $\nu'=0.432$.
\label{tau_alpha}}
\end{figure}

In the absence of  obvious symmetry breaking, it is useful to characterize the dynamics using van Hove  correlation function to discern the onset of glass-like behavior \cite{kirkpatrick1988PRA}.
The correlation function,  
\begin{equation}
F_{\vec{q}} (t) = \frac{1}{N} \sum_{j=1} ^N e^{i\vec{q}\cdot\left(\vec{r}_j(t)-\vec{r}_j(0)\right)},
\end{equation}
provides dynamical information of how the system relaxes from its initial configuration,
where $\vec{r}_j(t)$ is a position of $j^{th}$ monomer at time $t$.
The ensemble-averaged isotropic self-intermediate scattering function $\langle F_q (t) \rangle$ is estimated by integrating $F_{\vec{q}} (t)$ over space with $q = |\vec{q}|$ and at $q = q_{\text{max}} = 2\pi/r_s$, where $r_s$ is the position of the first peak in the total pair distribution function (see Fig.\ref{compaction}c). 
The onset of the structural glass transition is described by the density-density correlation function $\langle F_q (t)\rangle$ as a natural order parameter, which decays to zero in the liquid phase, but saturates to a non-zero value in the glassy phase even at long times. 
Thus, $\langle F_{q_{\text{max}}} (t) \rangle$ provides information of how rapidly the polymer confined to a sphere loses  memory of the initial configuration (Fig.\ref{tau_alpha}a). 
From physical considerations, $\langle F_{q_{\text{max}}} (t) \rangle$ should vanish at long times ($t\rightarrow\infty$) for $\phi < \phi_c$; 
the decorrelation time of the polymer configuration increases sharply as the extent of confinement (or $\phi$) approaches its dynamical arrest value.  
$\langle F_{q_{\text{max}}} (t) \rangle$ at various $\phi$ is well fit by a stretched exponential function $\sim e^{-(t/\tau_{\alpha})^{\beta}}$, and the dependence of $\tau_{\alpha}$ on $\phi$ for different $N$ (Fig.\ref{tau_alpha}) is analyzed using the relation, 
\begin{equation}
\tau_{\alpha}(\phi;N)=\tau_0(N)(\phi_c(N)-\phi)^{-\nu_{\tau}}.
\label{eqn:tau_scaling} 
\end{equation}  
The relaxation time $\tau_{\alpha}(\phi;N)$ increases with $\phi$ and diverges at $\phi_c(N)$. The stretching exponent $\beta$ decreases with $\phi$ (Fig.S4), in consistent with our findings in Fig.\ref{compaction}a that the system becomes more glassy as $\phi$ increases.  
The set of $\tau_{\alpha}(\phi;N)$, for various $N$, are described by a universal curve, satisfying $\log{(\tau_{\alpha}/\tau_0)}=-\nu_{\tau}\log{(\phi_c(N)-\phi)}$, and hence we obtain a universal scaling exponent $\nu_{\tau}\approx 0.65$ for the dynamical arrest.
The critical volume fraction $\phi_c(N)$ is $N$-dependent but saturates to a finite value $\phi_c^{\infty}$ in the limit $N\rightarrow\infty$. 
From finite size scaling (Fig.\ref{tau_alpha}c), we obtain $\phi_c^{\infty}\approx 0.45$. 

\begin{figure}
\includegraphics[width=1.0\columnwidth]{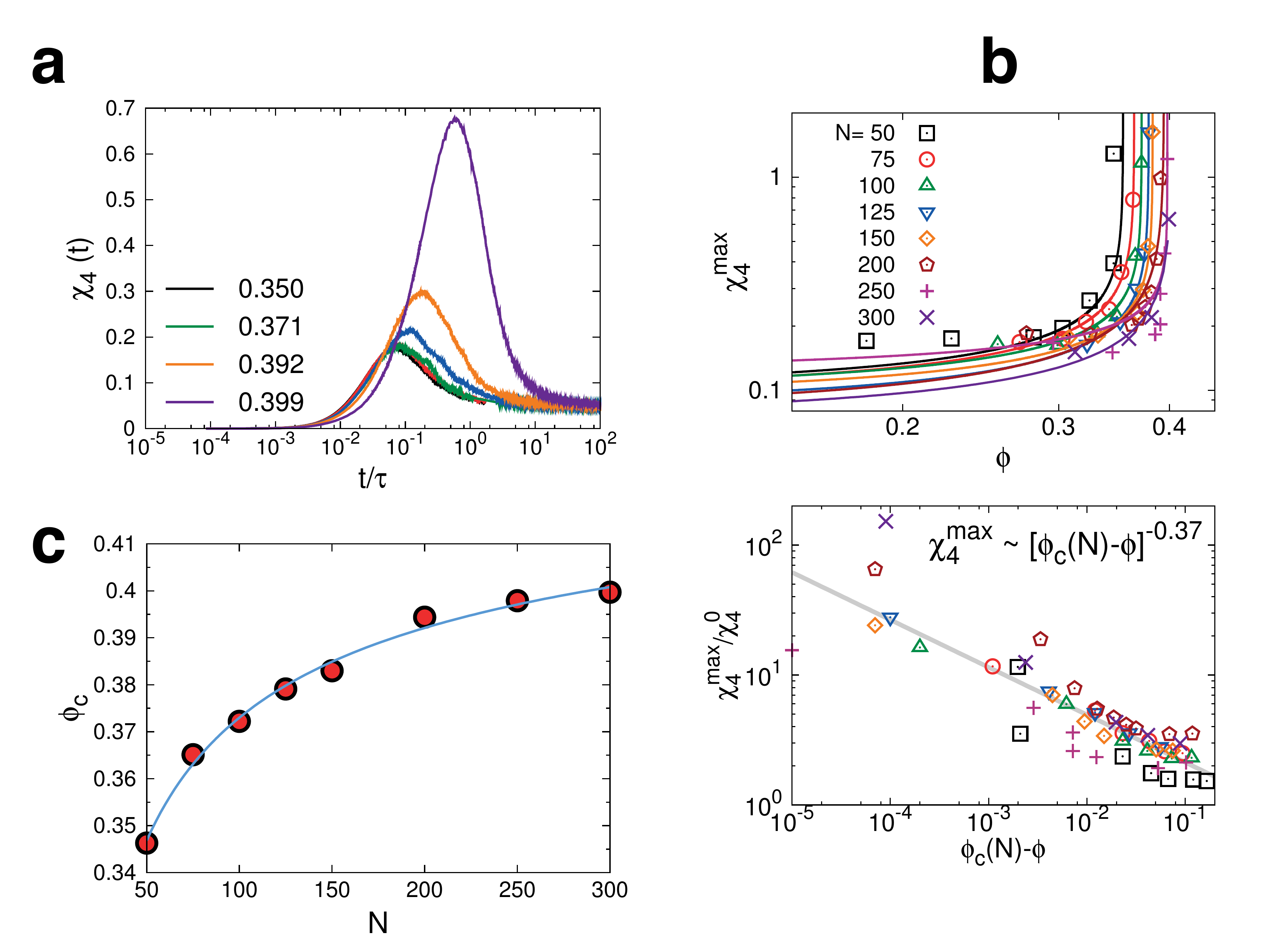}
\caption{
Polymer dynamics under confinement probed using $\chi_4^{\text{max}}(\phi)$.
(a) $\chi_4^{\text{max}}(t)$ with varying $\phi$ for $N=300$. 
(b) (top) $\chi_4^{\text{max}}(\phi)=\chi^o_4(N)(\phi_c(N)-\phi)^{-\nu_{\chi}}$ diverges at $N$-dependent critical volume fraction $\phi_c(N)$.
(bottom) The data of $\tau_{\alpha}$ are rescaled with $\tau_0(N)$ and the fit  using $\chi_4^{\text{max}}/\chi_4^o(N)=(\phi_c(N)-\phi)^{-\nu_{\chi}}$ for all $N$, which gives $\nu_{\chi}\approx 0.37 $. 
(c) Finite size scaling: $\phi_c(N)$ fit to
$\phi_c(N)=\phi_c^{\infty}-aN^{-\nu''}$ gives 
$\phi_c^{\infty}=0.443$, $a=0.571$, and $\nu''=0.453$. Note that $\nu'\approx \nu''$ (Fig. 2c). 
\label{chi4}}
\end{figure}

As an alternative to $\tau_{\alpha}(\phi)$, the fluctuations in $F_{q_{\text{max}}}(t)$, namely the generalized susceptibility $\chi_4(t)$ corresponding to the variance in $F_{q_{\text{max}}}(t)$, can distinguish between the states below and above $\phi_c$ clearly. The fourth order dynamic susceptibility \cite{kirkpatrick1988PRA}, used to quantify dynamic heterogeneity in structural glasses, is given by 
\begin{equation}
\chi_4 (t) = N\left[\langle F_{q_{\text{max}}}(t)^2 \rangle - \langle F_{q_{\text{max}}}(t) \rangle^2\right].
\end{equation}
The amplitude of $\chi_4 (t)$, $\chi^{\text{max}}_4$, increases with $\phi$ (Fig.\ref{chi4}a), and  
the divergence of $\chi^{\text{max}}_4$ near $\phi_c(N)$ can be described using $\chi_4^{\text{max}}(\phi;N)=\chi_4^o(N)(\phi_c(N)-\phi)^{-\nu_{\chi}}$. 
The scaling exponent for dynamical arrest transition is found to be $\nu_{\chi}\approx 0.37$.
In the $N\rightarrow\infty$ limit, $\phi_c^{\infty}=0.44$ (Fig.\ref{chi4}c), which is consistent with the $\phi_c^{\infty}=0.45$ from the analysis  based on Eq.\ref{eqn:tau_scaling}. 

The significance of the key finding that $\phi_c^{\infty}\approx 0.44$ becomes transparent by predicting the consequences for chromosome dynamics in various organisms.
Without confinement or any special interactions mediated by proteins,
genome occupies a large volume $V\sim \frac{4\pi}{3} (R_g^o)^3$ with $R_g^o\approx l_p(N/g)^{3/5}$ ($l_p\approx 50$ nm $=g\times 0.34$ nm/bp, thus $g\approx 147$ bp). 
Given that the nuclear sizes are similar ($\sim\mathcal{O}(1)$ $\mu m$), there could be a large variation in the nuclear volume fraction for different organisms that have different genomic size, $N$. 

(i) For bacteria ($N=10^6$ bp), $R_g^o\approx 10$ $\mu$m  is greater than the bacterial cell size $\sim 1$ $\mu$m. As 1 bp corresponds to 1 nm$^3$ \cite{phillips2009physical}, the volume fraction for bacterial genome is
$\phi_{\text{bac}}=1 \text{ nm}^3/\text{bp}\times10^6 \text{ bp}/1\mu m^3=10^{-3}\ll \phi_c^{\infty}$, implying that glassy effects are not relevant.

(ii) 
In eukaryotes, DNA chains are organized in nucleosomes.  
Thus, it is more appropriate to estimate the volume of chromosomes in terms of the number of nucleosomes rather than volume of bare DNA. 
Since each nucleosome, whose volume is $V_{\text{nuc}}\approx  \pi \times 10^2 \times 3$ nm$^3$ (15--20 nm width, 2--3 nm height), is   
wrapped by $\sim$150 bp DNA with a 50 bp-spacer between the neighboring nucleosomes \cite{hui1999CellResearch}, 200 bp-DNA is required to compose one nucleosome. 
For budding yeast $N\approx 10^8$ \cite{goffeau1996Science}, 
the volume occupied by the entire nucleosomes is $V^{\text{yeast}}_{\text{gen}}=10^8\text{ bp}/200\text{ bp}\times V_{\text{nuc}}\approx 0.47$ $\mu$m$^3$; and the yeast nucleus volume is  $V^{\text{yeast}}_{\text{nucls}}\approx 4$ $\mu$m$^3$. 
Therefore, $\phi_{\text{yeast}}=V^{\text{yeast}}_{\text{gen}}/V^{\text{yeast}}_{\text{nucls}}\approx 0.12$, which is smaller than $\phi_c^{\infty}$.  
This explains the intrachromosomal contact frequency $P(s)\sim s^{-1.5}$ for yeast genome, pointing to an equilibrium globule \cite{Wong12CurrBiol,emanuel2009PhysBiol,Therizols2010PNAS,Mirny11ChromoRes}.

(iii) 
Human nucleus size varies depending on the cell type and the stage of development, which results in $P(s)\sim s^{-1.5}$ for stem cell and $P(s)\sim s^{-1.0}$ for mature cell \cite{Barbieri12PNAS}.    
For illustrative purposes, we adopt the nucleus volume $V^{\text{human}}_{\text{nucls}}\approx 60-110$ $\mu$m$^3$ from the average size of mammalian cell nucleus $2\times R \approx 5-6$ $\mu$m \cite{Mirny11ChromoRes,AlbertsBook}. 
Since the volume taken by the entire 46 chromosomes, as a diploid with $2\times N\approx 2\times 3 \times 10^9$ bp, is
$V^{\text{human}}_{\text{gen}}=6\times 10^9\text{ bp}/200\text{ bp}\times V_{\text{nuc}}\approx 3\times 10^{10}$ nm$^3$, 
the volume fraction of human genome is $\phi_{\text{human}}=V^{\text{human}}_{\text{gen}}/V^{\text{human}}_{\text{nucls}}\approx 0.3-0.5\gtrsim \phi_c^{\infty}$.
Of particular note is that $\phi_{\text{human}}\gtrsim \phi_c^{\infty}$. 
Thus, the lack of nuclear space in human cell makes the chromosome dynamics intrinsically glassy, indefinitely slowing down the relaxation of chromosome configuration. 
This crucial conclusion based on the simple estimate of $\phi_{\text{human}}$ suggests the decondensation-condensation process, driven by a panoply of partner enzymes, is likely to be under kinetic control. 

(iv) The volume fraction of DNA ($L\approx 6-60$ $\mu$m) inside a viral capsid ($R\approx 25-50$ nm) using $L\sim 30$ $\mu$m and $R\sim 35$ nm is  $\phi_{\text{virus}}\approx 0.5>\phi_c$ \cite{purohit2005BJ,Halverson14RPP}.
A recent experiment showed that dynamics of viral packaging is ultraslow and glassy resulting in significant heterogeneity in packaging rates that vary from one virus to another \cite{Berndsen14PNAS}. 
It is noteworthy that the size of viral DNA ($N=(30\times 10^3)/0.34=8.8\times 10^4$ bp) is only $\sim$10 \% of bacterial genome.  Thus, the equilibration time of DNA conformation based on reptation (or scaling) should occur $10^3$ times faster than  in bacterial genome, which would contradict experiments \cite{Berndsen14PNAS}. To explain the ultraslow and heterogeneous dynamics of viral DNA packaging it is essential to consider the effects of confinement, and
our theory provides a natural explanation of the observations.  

Our study provides a general framework to quantify glassy dynamics of a polymer chain 
(a simple model for chromosome organization) 
and highlights the non-equilibrium aspect of a single polymer under strong confinement with clear implications for the variations in genome folding across different species. 
Dynamical implication of our finding $\phi_{\text{bac}}\ll\phi_{\text{yeast}}<\phi_c^{\infty}\lesssim\phi_{\text{human}}<\phi_{\text{virus}}$, and the correlations of $\phi_{\text{yeast}}<\phi_c^{\infty}$ with $P(s)\sim s^{-1.5}$ for budding yeast \cite{Wong12CurrBiol} and $\phi_c^{\infty}\lesssim\phi_{\text{human}}$  with $P(s)\sim s^{-1}$ for mature human cells \cite{Barbieri12PNAS} provide a new framework for understanding  the origin of qualitatively distinct chromosome organization in various organisms and cell types. 

Given that cellular environment is replete with crowding particles, the volume fractions estimated here for different organisms may well be only lower bounds, and thus we expect that glassy dynamics is prevalent especially in higher-order organisms. 
To overcome  topological constraints, fluidization or equilibration of nuclear environment using topoisomerase or metabolic activity would be sometimes necessary for biological systems to execute their functions  \cite{parry2014Cell}.
Furthermore, it is noteworthy that although there is not significant difference in genome volume fraction between human embryonic stem cell (hESC) and mature cell \cite{Pagliara2014NatureMaterials}, these two cells have distinct $P(s)$ ($P(s)\sim s^{-1.5}$ for hESC, $P(s)\sim s^{-1.0}$ for mature cell) \cite{Barbieri12PNAS,Zhang15PNAS}, which may be linked to substantial variations in metabolic activity or specific interactions with nuclear envelope depending on the cell maturity.
Although our conclusions here do not consider the role of active mechanisms on genome organization, it is plausible that equilibration machineries exploiting active forces are required when chromosome dynamics is intrinsically glassy, as appears to be the case in higher organisms. \\

\begin{acknowledgments}
We thank Pavel Zhuravlev for useful comments. This work was supported in part by a grant from the National Science Foundation (CHE 13-61946) (D.T.). C.H. thanks the KITP at the University of California, Santa Barbara (Grant No. NSF PHY11-25915), for support during the preparation of the manuscript.
We thank CAC in KIAS and KISTI for supercomputing resources (KSC-2014-C1-036).
\end{acknowledgments}

\bibliography{mybib1}

\setcounter{figure}{0}
\makeatletter 
\renewcommand{\thefigure}{S\@arabic\c@figure}
\makeatother 

\section*{Supplemental Material}
{\bf Model. }
In order to assess the conditions describing the onset of glassy dynamics of a confined flexible polymer we introduce a model in which the potential energy is given by
\begin{align}
U(\vec{r}_1,...,\vec{r}_N) &=\sum_{i=1}^{N-1}U^{\text{bond}}_i+\sum_{i = 1} ^{N-1} \sum_{j = i+1}^NU^{\text{ex}}_{i,j}+\sum_{i=1}^NU^{\text{surf}}_i\nonumber\\
&= \frac{k}{2} \sum_{i = 1} ^{N-1} \frac{(|\vec{r}_{i+1} - \vec{r}_i | - a)^2}{a^2} \nonumber\\
&+ \sum_{i = 1} ^{N-1} \sum_{j = i+1}^N \epsilon \left(\frac{a}{|\vec{r}_i - \vec{r}_j|}\right)^{12} \nonumber\\
&+ \sum_{i = 1} ^{N} 4 \epsilon \left(\frac{a}{R_s - |\vec{r}_i|}\right)^{12},
\end{align}
where $a$ is a bond length, $k = 80 \thinspace k_{\rm B} T$, $\epsilon = 2 \thinspace k_{\rm B} T$, $\vec{r}_i$ is the position of the $i^{th}$ monomer. 
To model the effect of confinement we placed the polymer chain in a sphere surface of radius $R_s$. 
The interaction between the monomers and the sphere surface is repulsive, given by the last term in Eq.(1).

We performed Brownian dynamics simulations of a self-avoiding polymer under spherical confinement with $N=$ 50, 75, 100, 125, 150, 200, 250, and 300 by integrating the following equations of motion,
\begin{equation}
\zeta \frac{d\vec{r}_i}{dt}=-\nabla_{\vec{r}_i}U(\vec{r}_1, \cdots, \vec{r}_N) + \vec{\Gamma}_i(t),
\end{equation}
where $\vec{\Gamma}_i(t)$ is the Gaussian random force satisfying the fluctuation-dissipation theorem, 
$\langle \vec{\Gamma}_i(t)\cdot\vec{\Gamma}_j(t^\prime) \rangle = 6k_{\rm B} T \zeta \delta(t-t^\prime) \delta_{ij}$.
With the Brownian time defined as $\tau = a^2/D$ where $D=k_BT/\zeta$, we chose the integration time step $\delta t=8.6\times 10^{-6}$ $\tau$ as a compromise between accuracy and computational cost. 

For eukaryotic genome, $D=\frac{k_BT}{6\pi\eta(a/2)}=\frac{4.14 pN\cdot nm}{6\pi\times (0.89\times 10^{-3}N/m^2\cdot sec)\times 10 nm} \approx 250$ $\mu m^2/s$ with $\eta = 0.89 \times 10^{-3} Pa \cdot s$ and $a\approx 20 $ nm; and hence we set $\tau = a^2/D\approx 1.6$ $\mu s$ and $\delta t\approx 13.7$ ps. 

We gradually reduced $R_s$ from $4R_g$ to the value at which $\tau_{\alpha}$ or $\chi^{\text{max}}_4$ (see Eqs. (2) and (3) in the main text) starts to diverge.
Although the detailed procedure of reducing the confinement size $R_{s}$ varies with $N$, the rate of $R_s$ reduction $r=\Delta R_s/\Delta t\sim 0.15$ $(R_s/a)/(2\times 10^8\times \delta t)$ is almost identical for all $N$ when $R_s$ approaches to the point of dynamical arrest.  
The $R_{s}$ values varied in the simulations are listed in Table \ref{table:Rs}, and the time-dependent protocol of reducing $R_s$ is plotted in Fig.~S1.
At each $R_{s}^{(i)}$, we simulated for $2\times 10^8$ $\delta t$ and took the last conformation from the previous simulation at $R_{s}^{(i-1)}$ as the initial conformation for simulation in $R_s^{(i)}$. 
We reduced $R_{s}$ from $R_{s}^{(i-1)}$ to $R_s^{(i)}$ linearly for $2\times10^{4}\delta t$, allocated the next $2\times 10^6$ $\delta t$ for an equilibration, and used the rest of $2\times 10^8$ steps to calculate $F_{q_{\text{max}}}(t)$ and $\chi_4^{\text{max}}$.   
We generated 10 independent trajectories for $N\leq150$ and 25 for $N\geq200$ to improve the quality of statistics.

It is worth emphasizing that the critical volume fraction $\phi_c$ is robust and insensitive to the range of confining speed. 
To show this, we used two different confining speeds for a polymer with $N=150$: one is $r_f=-0.30(R_s/a)/{\text{step}}$, 2 times faster than $r$ and the other is $-0.10(R_s/a)/{\text{step}}$, 1.5 times slower than $r$. 
Both from $\tau_\alpha$ using $F_q(t)$ and $\chi_{4}^{\text{max}}$, we obtained $\phi_{c}=0.383$ for both quenched and annealed cases, which is in full agreement with the regular case (see Fig.\ref{speed}).
\\

\begin{table}
\begin{ruledtabular}
\begin{tabular}{ccccccccc}
$N$&50&75&100&125&150&200&250&300\\
\hline
$i=1$&$R_s=10a$&$10a$&$10a$&$10a$&$10.8a$&$13.3a$&$13.3a$&$13.3a$\\
$2$&$9.2a$&$9.2a$&$9.2a$&$9.2a$&$9.2a$&$11.7a$&$11.7a$&$11.7a$\\
$3$&$8.3a$&$8.3a$&$8.3a$&$8.3a$&$8.3a$&$10.8a$&$10.8a$&$10.8a$\\
$4$&$7.5a$&$7.5a$&$7.5a$&$7.5a$&$7.5a$&$9.2a$&$9.2a$&$9.2a$\\
$5$&$6.7a$&$6.7a$&$6.7a$&$6.7a$&$6.7a$&$8.3a$&$8.3a$&$8.3a$\\
$6$&$5.8a$&$5.8a$&$6.2a$&$6.2a$&$6.2a$&$7.5a$&$7.5a$&$7.5a$\\
$7$&$4.3a$&$4.3a$&$5.8a$&$5.8a$&$5.8a$&$6.7a$&$6.7a$&$6.7a$\\
$8$&$4a$&$4a$&$5.5a$&$5.5a$&$5.5a$&$6.3a$&$6.3a$&$6.3a$\\
$9$&$3.7a$&$3.7a$&$5a$&$5a$&$5a$&$6a$&$6a$&$6a$\\
$10$&$3.3a$&$3.3a$&$4.7a$&$4.7a$&$4.7a$&$5.7a$&$5.7a$&$5.7a$\\
$11$&$3a$&$3a$&$4.3a$&$4.3a$&$4.3a$&$5.3a$&$5.3a$&$5.3a$\\
$12$&2.7a&2.7a&$4a$&$4a$&$4.2a$&$5a$&$5a$&$5a$\\
$13$&2.5a&2.5a&$3.7a$&$3.7a$&$3.8a$&$4.7a$&$4.7a$&$4.7a$\\
$14$&2.3a&2.3a&$3.3a$&3.3a&$3.3a$&$4.3a$&$4.3a$&$4.3a$\\
$15$&2.2a&-&$3a$&3a&$3.2a$&$4.2a$&$4.2a$&$4.2a$\\
$16$&2a&-&$2.7a$&2.8a&$3a$&$4a$&$4a$&$4a$\\
$17$&-&-&-&2.5a&-&$3.7a$&$3.7a$&$3.8a$\\
$18$&-&-&-&-&-&-&-&$3.7a$\\
\end{tabular}
\end{ruledtabular}
\caption{\label{table:Rs} $R_{s}$ values used for the simulations with various $N$. In the table, $a$ is the monomer-monomer distance. $R_{s}$ was sequentially reduced from $i=1$ to $i=i_{\rm max}$ according to the procedure described in the SI text. The initial conformation was taken from the last conformation of the previous run except $i=1$ where we generated unconstrained chain conformation. We set $R_{s}^{(0)}$ to four times the $R_g$ of the unconstrained chain, and slowly decreased $R^{(0)}_{s}$ to $R_{s}^{(1)}$.}
\end{table}

\begin{figure}
\includegraphics[width=0.8\columnwidth]{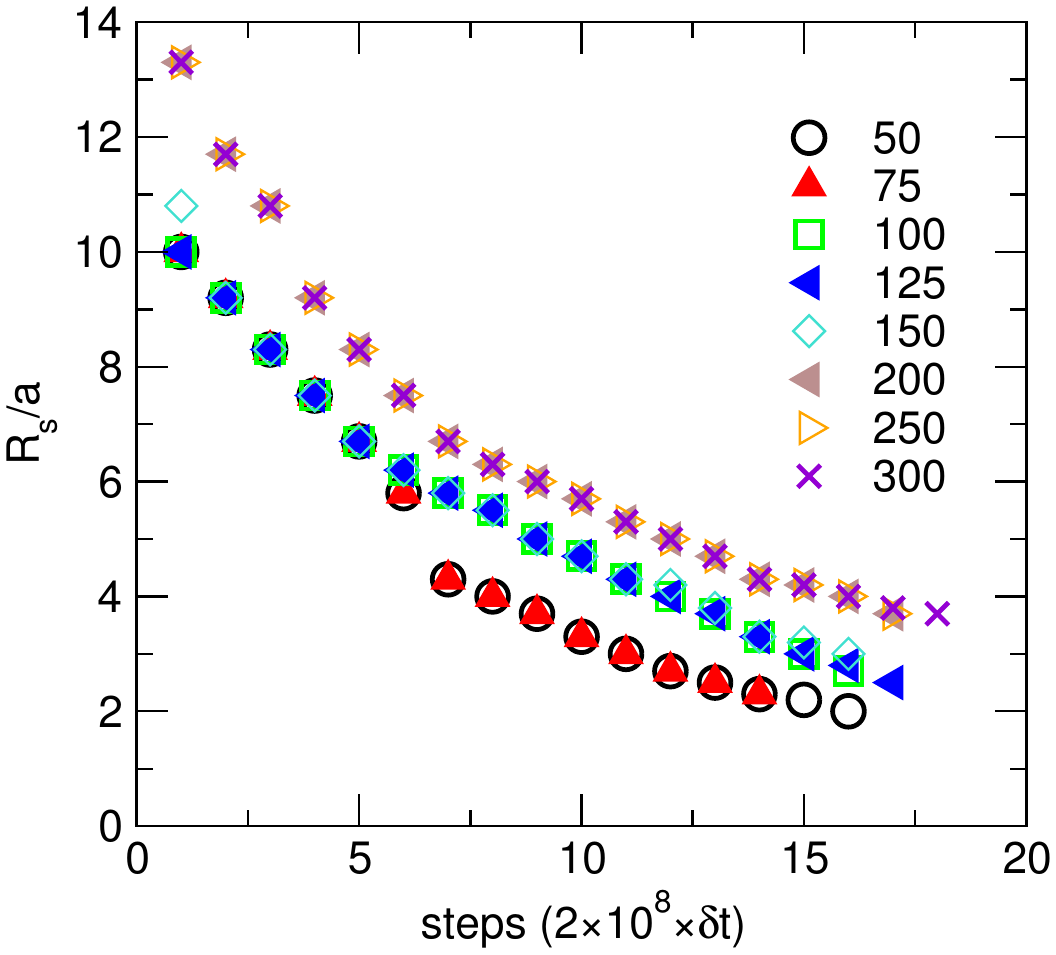}
\caption{The protocol used to reduce the confinement size ($R_s$) for different $N$. 
The reduction rate of confinement size near the dynamical arrest point is similar for all $N$ as $r\sim -0.15 (R_s/a)/\text{step}$. 
\label{chi4}}
\end{figure}

\begin{figure}
\includegraphics[width=1.0\columnwidth]{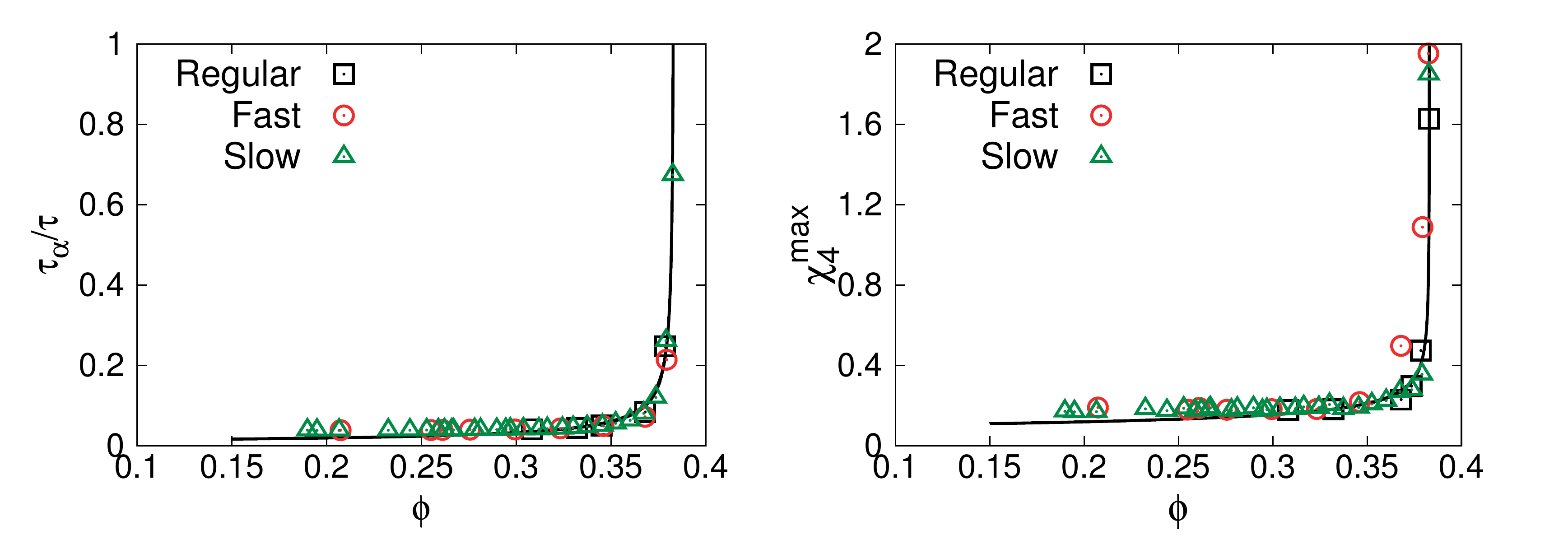}
\caption{Robustness of $\phi_c$ value for different reduction rate of confinement size. 
In the legend, ``Fast" denotes the faster confining speed; ``Slow" is for the slower one; and ``Regular" is the speed used in Table 1. 
Black lines are the fits for the values obtained at the regular speed. 
In all cases, we obtained $\phi_c=0.383$ for $N=150$. 
 \label{speed}}
\end{figure}

\begin{figure}
\includegraphics[width=1.0\columnwidth]{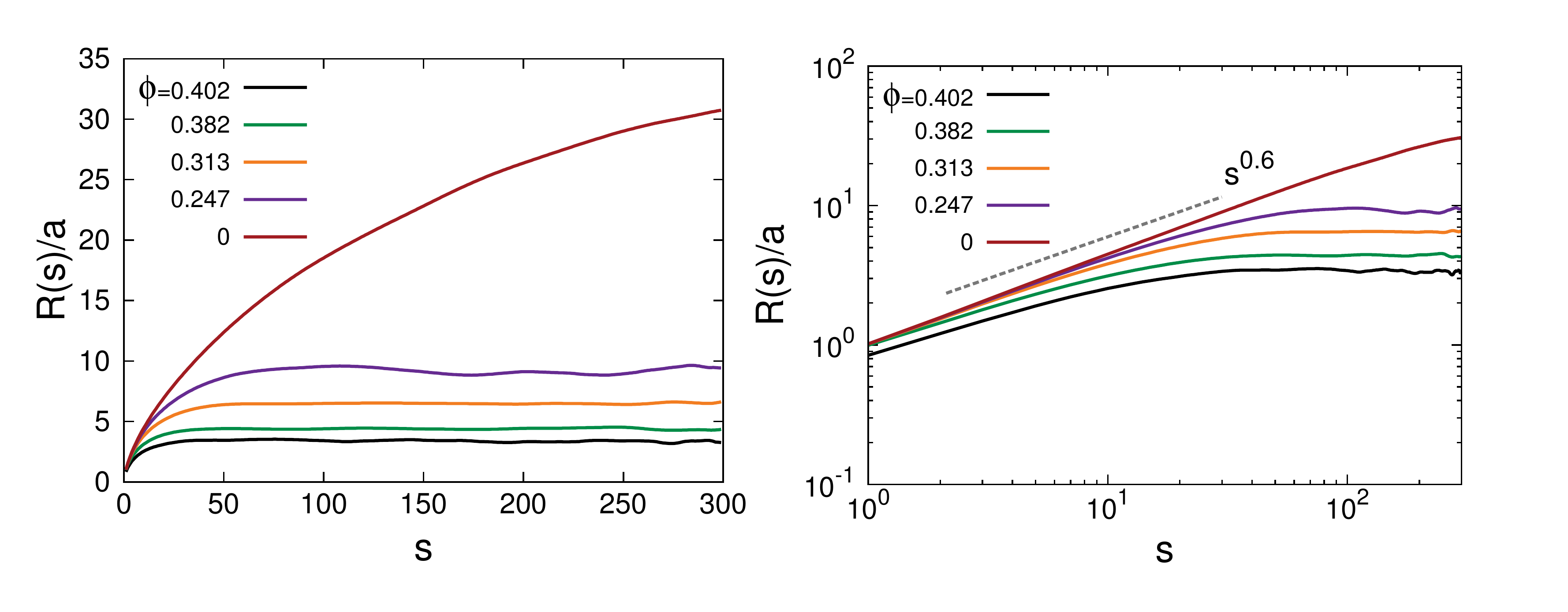}
\caption{Mean spatial distance of polymer with $N=300$ as a function of intersegmental separation $s$ for varying volume fraction $\phi$.  
Log-log plot is shown on the right panel with the dotted line expected for the scaling of SAW ($R(s)\sim s^{0.6}$).   
Note that the condition of confinement ($R_g^o>R_s$) trivially gives rise to the plateauing of $R(s)$ \cite{mateos2009PNAS,Barbieri12PNAS}.  
\label{Rs}}
\end{figure}

\begin{figure}
\includegraphics[width=0.7\columnwidth]{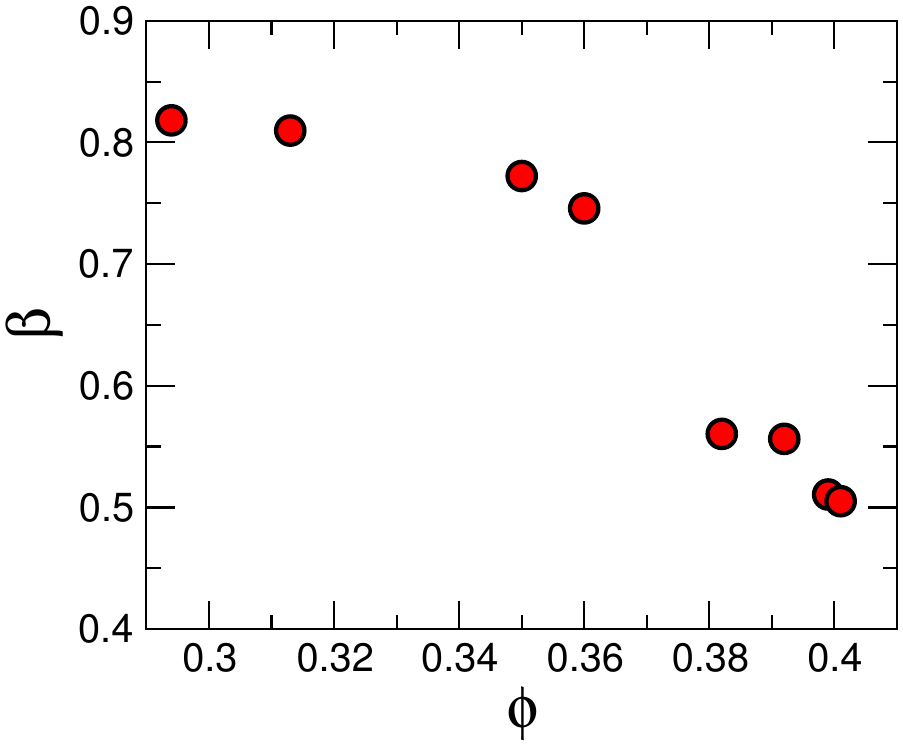}
\caption{$\beta$ value from the fit of $F_{q_{\text{max}}}(t)\sim e^{-(t/\tau_{\alpha})^{\beta}}$ for the confined polymer ($N=300$) as a function of $\phi$. It is noteworthy that the decrease of $\beta$, the phenomenological stretching exponent that characterizes the extent of glassiness, is consistent with our observation that the polymer dynamics becomes more glassy with increasing $\phi$. 
\label{beta}}
\end{figure}

{\bf Volume fraction of a confined polymer.}
When a polymer is confined to a sphere with radius $R_s$, 
the size of the polymer $R_g^c$ can be related to the radius of gyration for polymer in free space ($R^o_g$) via the following scaling relation with $x=R^o_g/R_s$:
\begin{equation}
R_g^c=R^o_gf(x).
\end{equation}
(i) Under  weak confinement ($x\ll1$), corresponding to large $R_s$, the chain statistics will be unaltered $R_g^c\sim R^o_g\sim N^{\nu}$ with $\nu=3/5$, and thus $f(x)\sim$ constant.
(ii) In contrast, a strong confinement ($x\gg 1$) induces polymer collapse, so that $R_g^c\sim N^{1/d}$ and $f(x)\sim x^p$.
From $N^{1/d}\sim N^{\nu}(N^{\nu}/R_s)^p$, the exponent $p$ ought to be $p=(d\nu)^{-1}-1$.
Therefore, substituting $R^o_g=aN^{\nu}$ where $a$ is the Kuhn length, one gets
$R_g^c=R_s(a/R_s)^{1/d\nu}N^{1/d}$.

A definition of polymer volume fraction ($\phi$) using the ratio between $R_g^c$ and $R_s$, $\phi=(R_g^c/R_s)^d$ gives distinct scaling of $\phi$ with $N$, depending on the strength of confinement:
\begin{equation}
\phi=\left(\frac{R_g^c}{R_s}\right)^d=\left\{
\begin{array}{lr}
\left(\frac{a}{R_s}\right)^dN^{\nu d} : (\text{weak, } R^o_g\ll R_s)\\
\left(\frac{a}{R_s}\right)^{1/\nu}N : (\text{strong, } R^o_g\gg R_s)
\end{array}
\right.
\label{eqn:phi}
\end{equation}
where $1/\nu=d$ for the case of strong confinement.
Note that this definition of $\phi$ is invariant under coarse-graining.
\\

{\bf Radial distribution function. } 
We used 
\begin{equation}
g(r) = \frac{2}{N(N-1)}\sum^{N-1}_{i=1}\sum^N_{j=i+1}\delta(|\vec{r}_i-\vec{r}_j|-r)
\end{equation}
to capture the extent of packing between monomers in Fig.1c.
\\

{\bf Contact probability.}
Contact probability as a function of genomic separation $|i-j|=s$ in Fig.1d is given by,
\begin{equation}
P(s)=\frac{\sum^{N-1}_{i=1}\sum^N_{j=i+1}\delta(|i-j|-s)\Theta(a-|\vec{r}_i-\vec{r}_j|)}{\sum^{N-1}_{i=1}\sum^N_{j=i+1}\delta(|i-j|-s)}
\end{equation}
where $\Theta(\ldots)$ is the Heaviside step function. $\Theta(x)=1$ for $x\geq 0$; otherwise $\Theta(x)=0$. 
\\

\begin{figure}[ht]
\includegraphics[width=1.0\columnwidth]{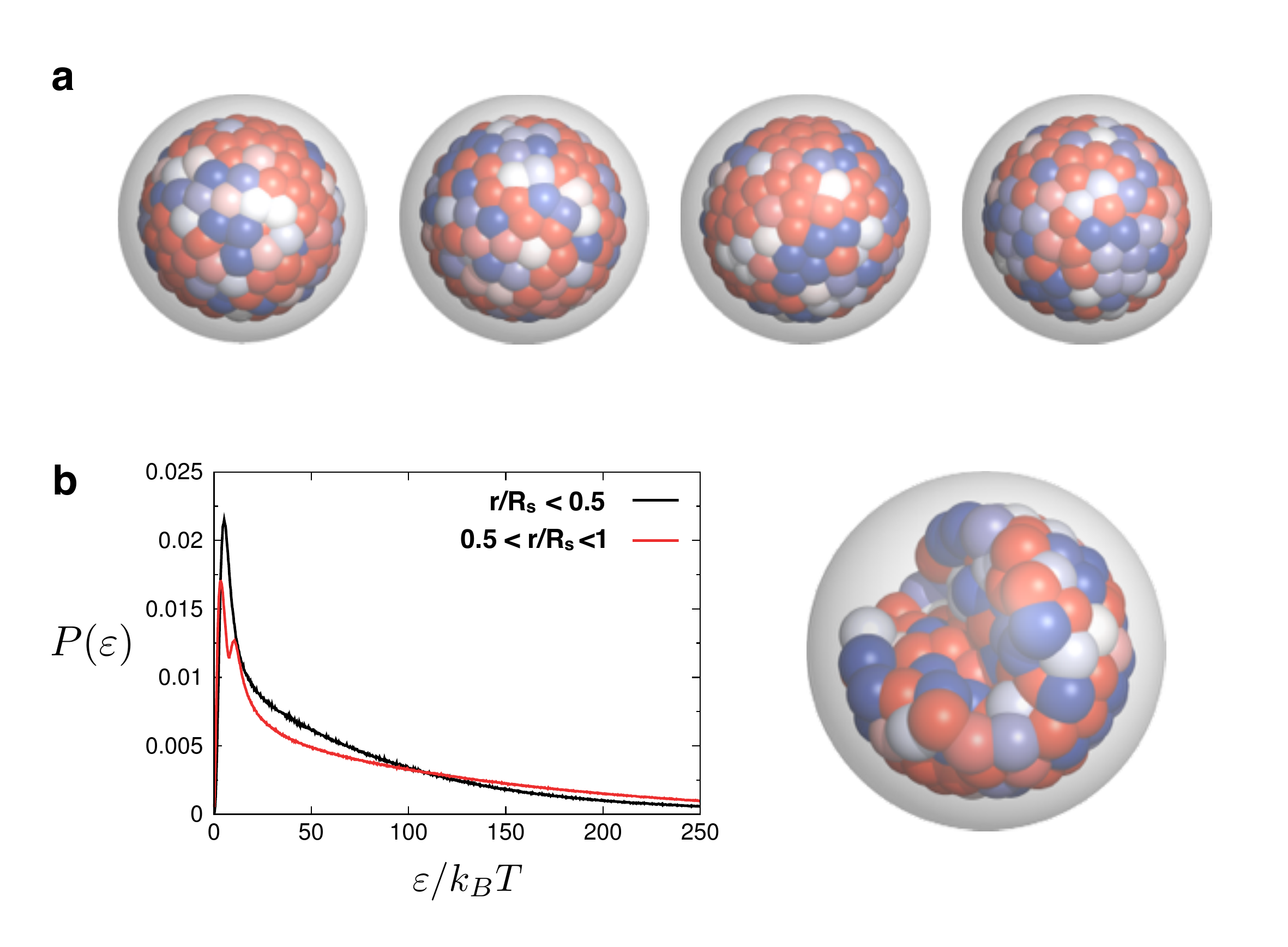}
\caption{(a) Snapshots of polymer under strong confinement ($\phi=0.402$). Monomers, colored based on the energy value, underscore the spatial heterogeneity of stress in the organization of the polymer.  
(b) Monomer energy distribution, $P(\varepsilon)$, at $\phi=0.402$ for different range of $r$: $r/R_s<0.5$ for core and $0.5<r/R_s<1$ for the surface. 
Together with the snapshot displaying the interior of the globule on the right, $P(\varepsilon)$ for the different range of $r$ highlights that the spatial heterogeneity of the monomer energy is present in the interior as well as on the surface of globule.  
\label{energy}}
\end{figure}

{\bf Scaling relationship of contact probability for SAW.}
In the absence of confinement, the chain statistics should obey that of self-avoiding walk. 
Given the distance distribution $P_s(r)$ between two interior points separated by $s$ along the contour, the contact probability is defined as $P(s)(\approx P_s(r=0))$.
From $P_s(r)\sim (1/s^{\nu})^df(r/s^{\nu})\sim (1/s^{\nu})^d(r/s^{\nu})^g$ for $r\ll s$, where $g$ is the correlation hole exponent and $g=\theta_2$ for two interior points \cite{desCloizeauxJP80}.
The scaling exponent should be similar to the probability of two interior points of a SAW chain to be in contact, $P(s)\sim s^{-(d+\theta_2)\nu}\approx s^{-2.18}$ with $d=3$, $\theta_2=0.71$, $\nu=0.588$ \cite{desCloizeauxJP80,Redner80JPA,Toan08JPCB}.
In accord with this expectation, our simulation shows $\alpha=2.18$ in the absence of confinement ($R_s/a\rightarrow\infty$). 
Note that for Gaussian chain (or polymer melt) $g=0$, $\nu=1/2$, and $d=3$, so that we retrieve the scaling relation for an equilibrium globule $P(s)\sim s^{-1.5}$ in the above.
\\

\begin{figure}[ht]
\includegraphics[width=1.0\columnwidth]{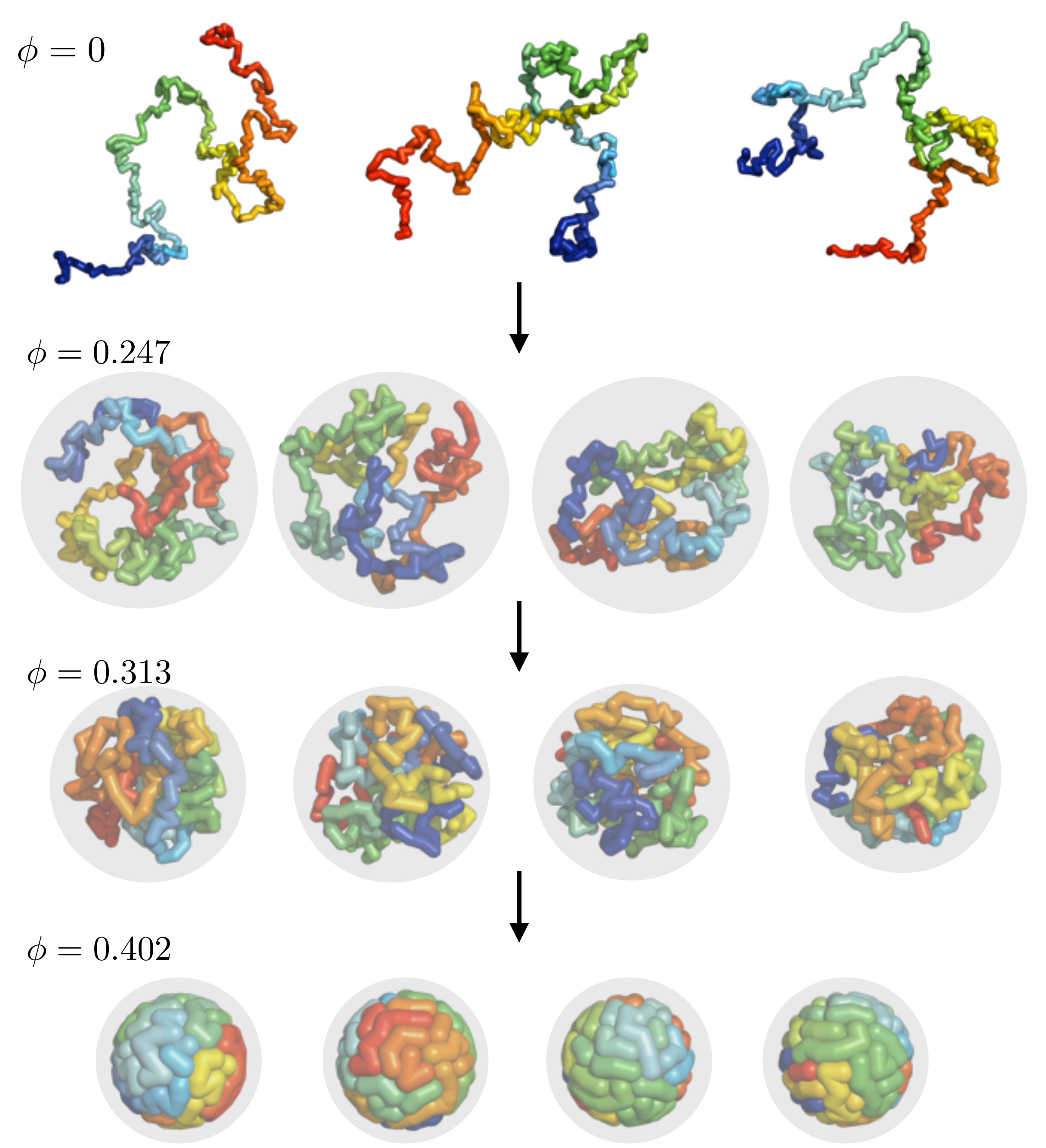}
\caption{Formation of fractal-like globules from self-avoiding chain with increasing extent of confinement ($\phi=0\rightarrow 0.402$). 
At $\phi=0.402$, the globules display segregated domains with ultra-slow mobility. \label{fractal_globules}}
\end{figure}

\end{document}